\documentclass{article}

\usepackage[T1]{fontenc}
\usepackage[utf8]{inputenc}
\usepackage{amsmath}
\usepackage{float}
\usepackage{url}
\usepackage{graphicx}
\usepackage{multirow}

\title{Extracting localized information from a Twitter corpus for flood prevention}
\author{Etienne Brangbour, Pierrick Bruneau, \\ Stéphane Marchand-Maillet, Renaud Hostache, \\ Patrick Matgen, Marco Chini, Thomas Tamisier}
\date{}

\begin{document}

\maketitle

\begin{abstract}
In this paper, we discuss the collection of a corpus associated to tropical storm Harvey, as well as its analysis from both spatial and topical perspectives. From the spatial perspective, our goal here is to get a first estimation of the quality and precision of the geographical information featured in the collected corpus. From a topical perspective, we discuss the representation of Twitter posts, and strategies to process an initially unlabeled corpus of tweets.
\end{abstract}

\section{Introduction}

Twitter is a popular micro-blogging service based on short textual posts. Its users interact by sharing their opinions, experiences or moods. Its availability on mobile platforms lets users react in real time to catastrophic events such as forest fires, earthquakes or floods. Modelling and predicting flood outbreak and spread are essential in order to mitigate the ecological and economical impact of such events. The classical approach for prediction relies on simulations of the flow and saturation resulting from rainwater \cite{bates00}. More recently, assimilation models were proposed as means to incorporate external information, such as multispectral satellite imagery, to ensembles of physical models \cite{hostache15}. The Publimape project explores the extension of this approach to social data, as posted on platforms such as Twitter or Instagram. The tropical storm Harvey, occurred in 2017 and widely discussed then\footnote{\url{https://en.wikipedia.org/wiki/Hurricane_Harvey}}, is used as a validation use case in the course of the project.

Using social data to perform event detection has been considered from two perspectives in the literature. From the one hand, \emph{Volunteered Geographical Information} explicitly asks users to go out on the field, and capture information linked to the event of interest \cite{griesbaum_direct_2017}. This point of view somehow relates to a crowdsourcing campaign. On the other hand, \emph{Participatory Sensing} views Twitter users as units of a sensor network \cite{burke_participatory_2006,crooks_earthquake:_2013}. Relevant information is then collected passively. The latter perspective is considered in the context of Publimape.

In this paper, we discuss the collection of a corpus associated to tropical storm Harvey, as well as its analysis from both spatial and topical perspectives. 
From the spatial perspective, our goal here is to get a first estimation of the quality and precision of the geographical information featured in the collected corpus. From a topical perspective, we discuss the representation of Twitter posts, and strategies to process an initially unlabeled corpus of tweets.

In section \ref{sec:sota}, we first review methodologies and constraints inherent to Twitter content collection in the context of event detection. The design choices for our use case is then described in section \ref{sec:harvey}. We also showcase insights derived from geographical information featured in Twitter content in section \ref{sec:experiments}. Adapted textual representations are discussed in section \ref{sec:textual}, and their categorization in section \ref{sec:classification}.

\section{Collecting Tweets for Event Detection} \label{sec:sota}

Twitter exposes an API that enables the real-time collection of posted messages\footnote{\url{https://developer.twitter.com/en/docs/tweets/filter-realtime/overview}}. It allows to filter the Twitter stream w.r.t. hashtags, keywords, or GPS bounding boxes. It is also possible to obtain a fully random sample of the stream, limited to 1\% of the complete stream rate of flow \cite{cheng_event_2014}. It is also possible to query content for a past period, but this is a paying feature (i.e. Twitter \emph{Enterprise} API\footnote{\url{https://developer.twitter.com/en/enterprise}}).

Tweets obtained via one of these APIs are collected as JSON objects, i.e. dictionary objects where fields may be literals (numbers or strings), or themselves dictionary objects, implementing a hierarchical data structure.

Among the root fields of a tweet object (approx. 30), in the context of this paper we will pay special attention to the following fields:

\begin{itemize}
\item \emph{coordinates}: the geotag (i.e. GPS coordinates) of the tweet,
\item \emph{place.full\_name}: the place attached to the tweet,
\item \emph{place.bounding\_box}: the bounding box of the above-mentioned place (\emph{bbox} in the remainder of the document),
\item \emph{user.location}: the place featured in the profile of the user that posted the tweet. Specifically, this is a free form text field.
\item \emph{text}: the actual tweet textual content. Whenever the \emph{truncated} field is true, \emph{text} is a truncated version, and the full version has to be fetched from \emph{extended\_tweet.full\_text}.
\end{itemize}

Sharing tweet databases is forbidden \emph{a priori}. However, sharing collections of tweet IDs is allowed, delegating the charge of \emph{hydrating} them to full objects to the recipient. In the context of non-commercial research projects, it is hence possible to share ID collections of unlimited size\footnote{\url{https://developer.twitter.com/en/developer-terms/agreement-and-policy.html}}.

For collecting tweets in the context of event detection, the usage of keywords has been contrasted to that of geographical bounding boxes \cite{ozdikis_survey_2017}. Filtering w.r.t. keywords or hashtags has been commonly used \cite{starbird_chatter_2010}, for example in the context of a 2011 earthquake \cite{crooks_earthquake:_2013}. Spatio-temporal distributions of the filtered results may then be estimated \cite{sriram_short_2010,helwig_analyzing_2015}.

In \cite{cheng_event_2014}, the authors claim that collecting according to keywords neglects diffusion effects, i.e. when users copy and paste tweet text without explicitly using the retweet mechanism. Even if experiments in \cite{sakaki_tweet_2013} show a moderate impact of this phenomenon, collecting according to a geographical filter completely alleviates this bias \cite{ozdikis_survey_2017}. Geographical filters are also used in \cite{gao_mapping_2018}, excluding retweets and non-English language afterwards. Let us note here that the free streaming Twitter API is limited w.r.t. the Entreprise API in that it is impossible to query tweets posted by user having their profile in the Region of Interest (RoI). Only content explicitly tied to the RoI (i.e. coordinates or place fields) is retrieved. Our usage of geographical filters is described in section \ref{sec:harvey}.

Initiatives have recently emerged in favor of persistently sharing tweet collections. Harvard Dataverse\footnote{\url{https://dataverse.harvard.edu/}} stores such sets as an incentive for reproducible research. Actually, a set of 35M tweet IDs associated to hurricanes Harvey and Irma can be found there \cite{littman_hurricanes_2017}. Those were collected using keyword and hashtag filters: it is hence somehow complementary to the corpus introduced in section \ref{sec:harvey}. The former corpus served as basis for a MediaEval task recently \cite{bischke_multimedia_nodate}, where task managers selected a subset of 10K tweets featuring a picture, and had those pictures annotated w.r.t. the presence of a passable or flooded street using the Figure Eight platform\footnote{\url{https://www.figure-eight.com}}.

\section{Creation of the Harvey corpus} \label{sec:harvey}

\begin{figure}
\begin{center}
 \includegraphics[width=0.8\textwidth, keepaspectratio]{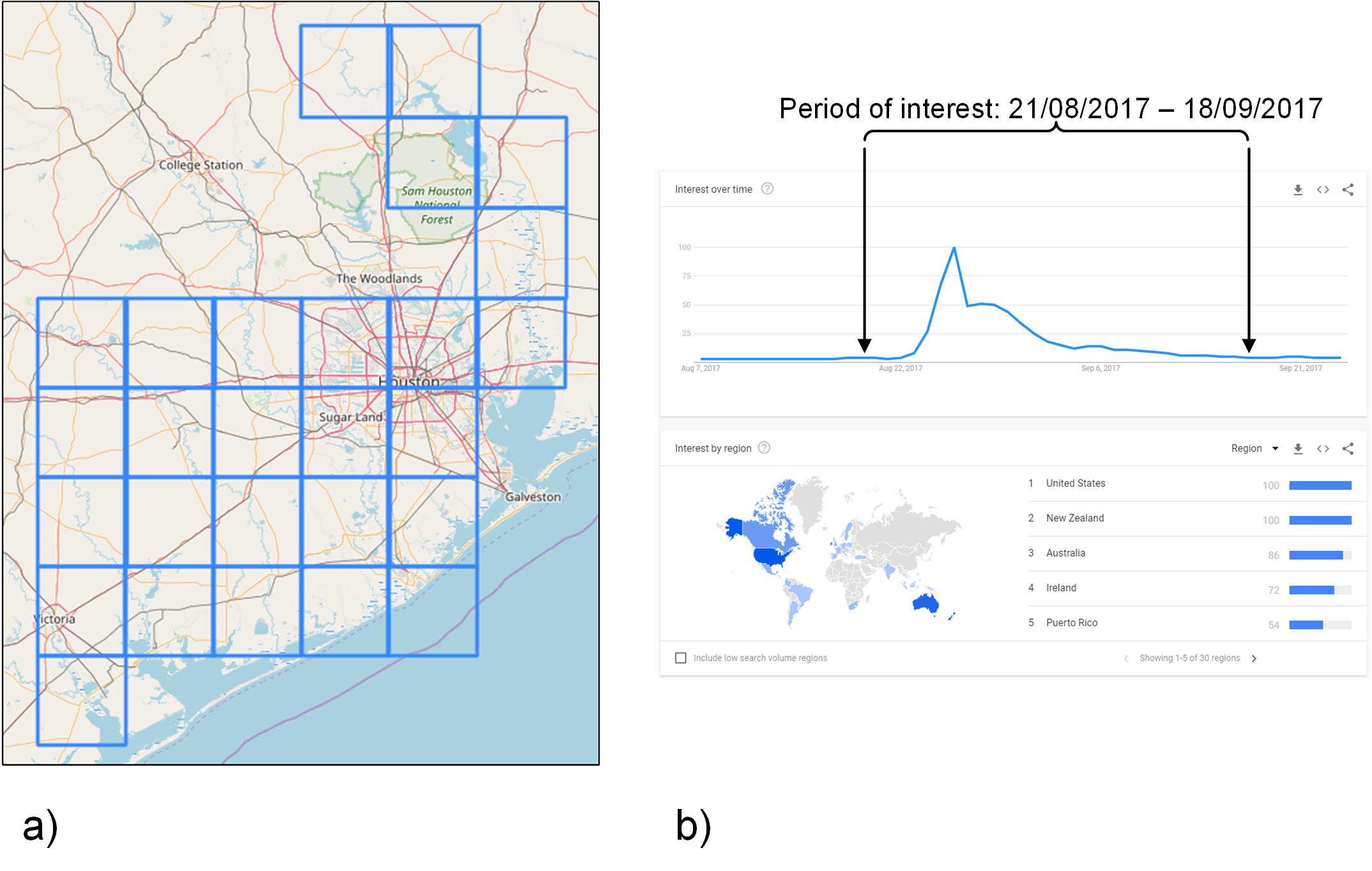}
 \caption{\emph{a)} RoI used for the corpus collection. \emph{b)} Popularity of the \emph{harvey} query through the period of interest according to \emph{Google Trends}.} 
 \label{fig:collection}
\end{center}
\end{figure}

With hydrologist partners in the project, we defined the RoI as indicated in figure \ref{fig:collection}a, for the period spanning from the 19$^{\text{th}}$ of August to the 21$^{\text{st}}$ of September. In figure \ref{fig:collection}b, we see this period is linked to a peak of queries featuring the term \emph{harvey}. 

Using such an ensemble of rectangular bounding boxes is required when using Twitter APIs. Using the Enterprise API, we query tweets with any of \emph{coordinates}, \emph{places}, or \emph{user.location} fields overlapping the RoI, for the period mentioned above. This lead to the collection of 7.5M tweets.

We store this data set as a document collection in a NoSQL database. In order to facilitate further spatial analysis, annotations of these documents are stored in a separate collection. An annotation is made of a tweet ID, and a type (\emph{geotag} or \emph{bbox}). For annotations of the \emph{geotag} type, GPS coordinates are attached directly as a field.  We also created a specific \emph{place} collection, that stores the detailed information for \emph{bbox} annotations (including the bounding box coordinates themselves). \emph{bbox} annotations refer to IDs from the \emph{place} collection as foreign keys.

\sloppy
Tweets obtained via the \emph{Enterprise} API have an extended user field (\emph{user.derived}), featuring normalized place names, but no associated bounding boxes, merely the \emph{geotag} a geographical centroid. As means to encode annotations of \emph{bbox} type, we used the Nominatim\footnote{\url{https://nominatim.openstreetmap.org/}} geocoding service to recover the bounding boxes. The risk of false recovery is limited by the fact that derived place names are obtained from Twitter's own geocoding service. We use the first retrieved bounding box that contains the geographical centroid.
We distinguish bounding boxes obtained from user profiles by creating annotations with type \emph{pbbox} (i.e. \emph{profile bounding box}). All types considered, we stored approximately 8.3M annotations, referring to 8434 distinct places.

Tweets were collected whether their \emph{coordinates}, \emph{place} or \emph{user} field overlaps the RoI defined in figure \ref{fig:collection}a. Let us note that tweets posted by a user profile matching the RoI, but with \emph{coordinates} or \emph{place} field out of the RoI is unlikely to be relevant for our use case (e.g. Houston inhabitant in vacation in Europe). Thus we post-process the annotation set, by excluding situations highlighted above. This led to exclude approximately 170K annotations and 4700 places, so approx. 2\% of annotations and more than 50\% of places. These numbers are consistent with the discussion above, and removes noise \emph{a priori} from the analysis presented in the next section.

\section{Geographical analysis of the corpus} \label{sec:experiments}

\begin{figure}
\begin{center}
 \includegraphics[width=0.95\textwidth, keepaspectratio]{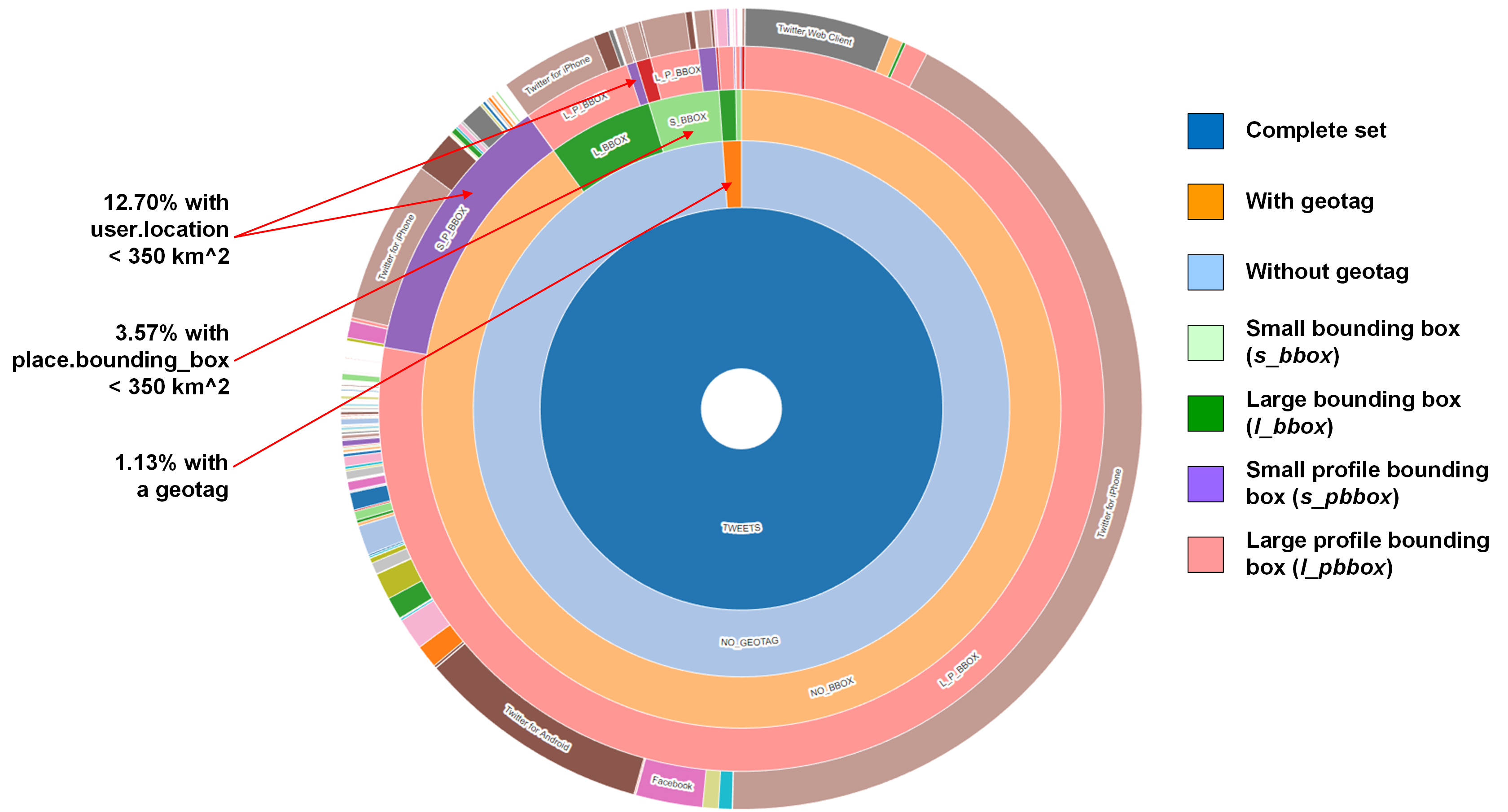}
 \caption{\emph{Sunburst} view of the annotation type cross distribution. The \emph{source} field is also reported.} 
 \label{fig:sunburst}
\end{center}
\end{figure}

One of the objectives of the project introduced in this paper is to map content posted on Twitter with as much spatial precision as possible. We initially assume that \emph{geotag} fields are exact. They make up for approximately 1\% of the annotation collection, as indicated in figure \ref{fig:sunburst}, and reported previously in the literature \cite{middleton_real-time_2014}.

\begin{figure}
\begin{center}
 \includegraphics[width=0.95\textwidth, keepaspectratio]{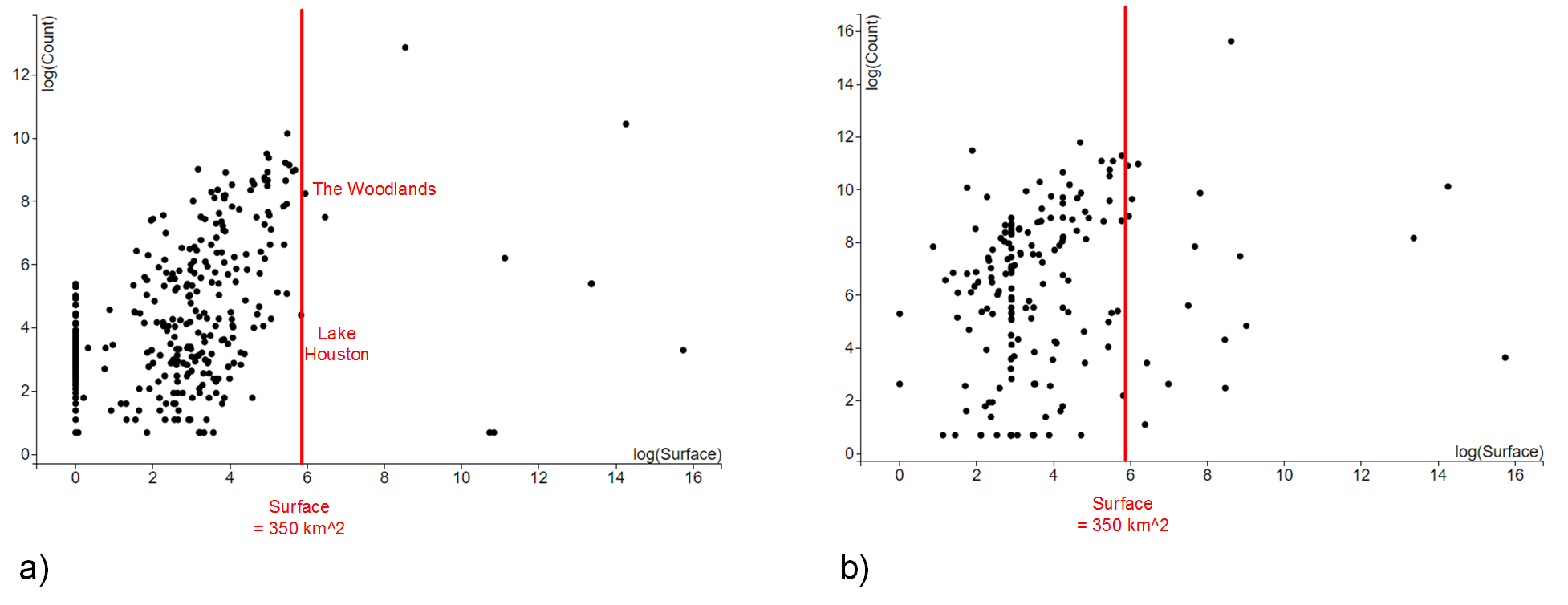}
 \caption{Scatterplots of the places frequency in the corpus as a function of their surface, for \emph{bbox} (\emph a) and \emph{pbbox} (\emph b) annotations.}
\label{fig:scatterplots}
\end{center}
\end{figure}

Hence we focus on the surfaces of the places identified in the corpus, specifically on the link between the surface of a bounding box and its frequency in the corpus: is a place with a large surface more frequent? Scatterplots in figure \ref{fig:scatterplots} display this link. A log-log scale is used for better legibility, as surfaces and frequencies, taken independently, are both exponentially distributed. Graphics in figure \ref{fig:scatterplots} were obtained from an interactive application where hovering over glyphs reveals the place name, its surface and its frequency, as means to facilitate exploration. \emph{bbox} and \emph{pbbox} annotations are represented separately.

\begin{figure}
\begin{center}
 \includegraphics[width=0.95\textwidth, keepaspectratio]{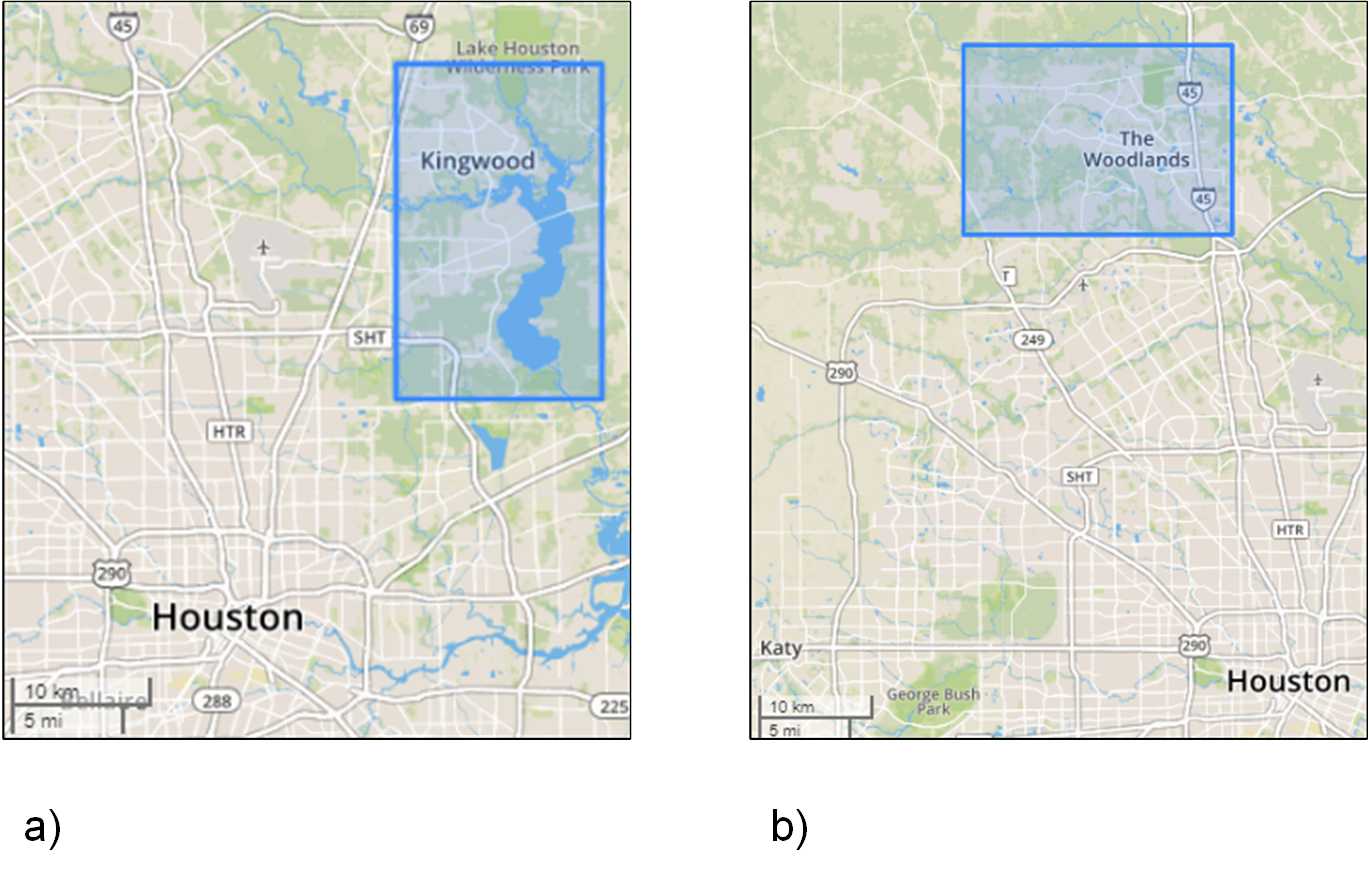}
 \caption{Bounding boxes for \emph{Lake Houston} (\emph a) and \emph{The Woodlands} (\emph b).} 
 \label{fig:threshold}
\end{center}
\end{figure}

For the \emph{bbox} case, a significant correlation exists according to Pearson's and Kendall's tests ($p < 10^{-10}$), with an estimation of $0.73$ for the Pearson correlation coefficient. Inspecting figure \ref{fig:scatterplots}a, we see that highly specific places like \emph{Cypress Park High School} are mentioned only 3 times, when \emph{Houston, TX} and \emph{Texas, USA} appear in respectively 3.9M and 34K tweets. For reliable social mapping, the latter places are not specific enough. In practice, we use figure \ref{fig:scatterplots}a to establish a specificity threshold beyond which a bounding box is not relevant for the use case of interest. We qualitatively set this threshold to 350 km$^2$, excluding points on the r.h.s. of \ref{fig:scatterplots}a. This value separates \emph{Lake Houston} from \emph{The Woodlands} (see figure \ref{fig:threshold}): this surface range is considerable, practically limiting the risk of excluding useful information from our study.

We transferred this threshold to the \emph{pbbox} annotations on figure \ref{fig:scatterplots}b. In the latter case, Pearson and Kendall are small ($0.19$ in both cases) and weakly significant ($p = 0.01$ for Pearson's test). Users often indicate a town as their profile location, which are generally less specific than locations given as examples before. However the chosen threshold retains most of them. 

We define sub-categories to \emph{bbox} and \emph{pbbox} according to our threshold (\emph{s} for \emph{small} and \emph{l} for \emph{large}). The cross distribution between annotation types is displayed in figure \ref{fig:sunburst}. There we see that only 17.4\% of the given geographical information is usable in our applicative context.

On figure \ref{fig:sunburst}, the \emph{source} field, indicating the originating application of the tweet, is also reported. The vast majority of tweets is emitted from iPhone, Android and web Twitter clients ($76\%$ for these 3 cumulated categories). Next most popular sources are other social platforms ($3\%$ for Facebook and $2\%$ for Instagram) and secondary Twitter clients ($1\%$ for both iPad and TweetDeck). Content posted by bots has minor presence in our corpus: the two first are SocialOomph and IFTTT with $1\%$ each.

Thanks to the interactive version of the view in figure \ref{fig:sunburst}, we focus on geographical information relevant to us (i.e. type \emph{geotag}, and the \emph{small} sub-types). Instagram is the origin of most tweets annotated by a \emph{geotag} ($63\%$). Focusing on \emph{geotag} annotations also reveals several minority resources relevant for our use case, e.g. meteorological (\emph{CWIS Twitter Feed}) and traffic (\emph{TTN HOU Traffic}) reports. However we still have to evaluate the reliability of geotags emitted by such automated resources.

A large majority of annotations in the \emph{s\_bbox} sub-category is sent from official Twitter clients. The latter also make up for 3 quarters of the \emph{s\_pbbox} sub-category. The last quarter gathers less popular clients mentioned above with many minor sources (e.g. most of IFTTT). Hence it is reasonable to first focus on \emph{geotag} and \emph{s\_pbbox} annotations.

\begin{figure}
\begin{center}
 \includegraphics[width=0.95\textwidth, keepaspectratio]{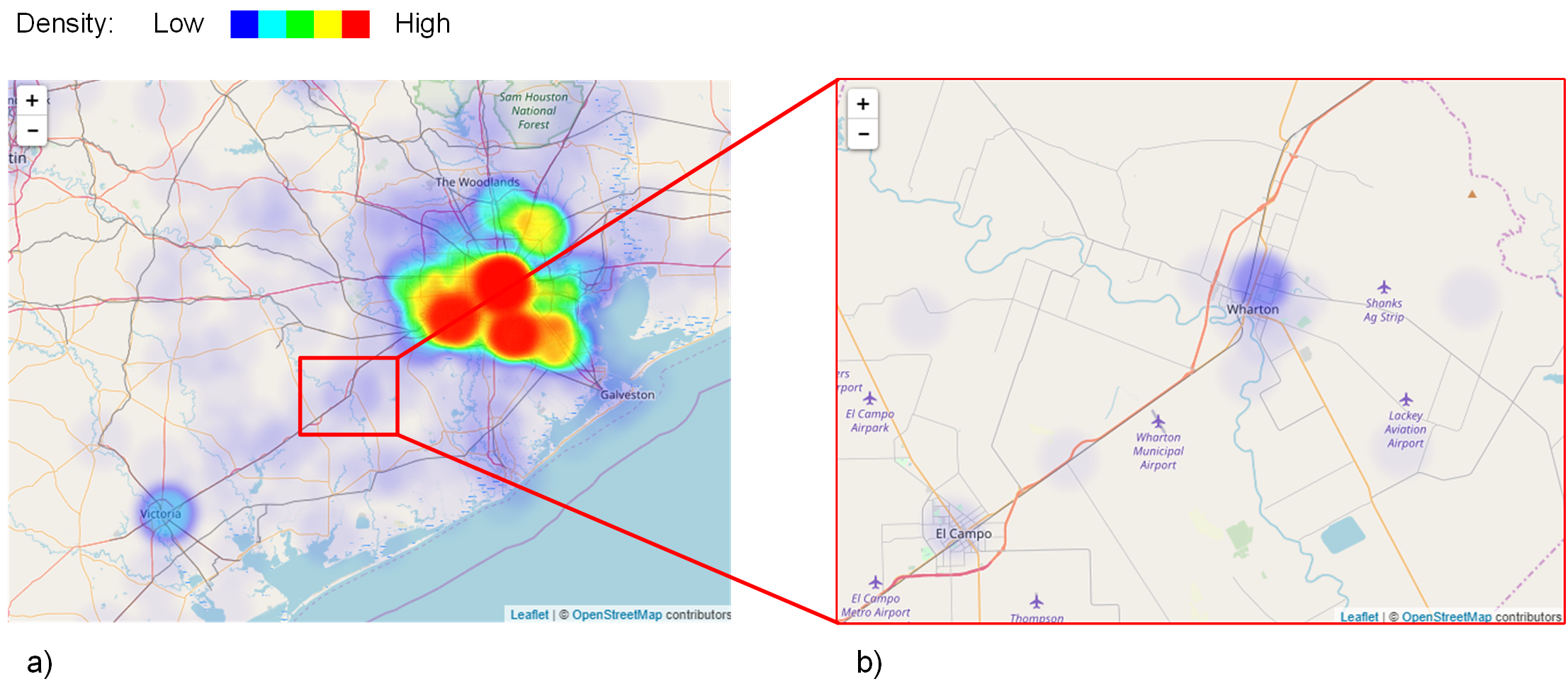}
 \caption{Density map of \emph{geotag} and \emph{s\_bbox} types among tweets featuring \emph{flood} or \emph{harvey} keywords (\emph a), zoomed on the Wharthon area (\emph b).} 
 \label{fig:heatmap}
\end{center}
\end{figure}

In order to facilitate discussions with hydrologist partners, among the \emph{geotag} and \emph{s\_bbox} categories, we filtered out tweets featuring \emph{flood} or \emph{harvey} in their text. We display the resulting 19K tweets on a density map (see figure \ref{fig:heatmap}a). Specifically, hydrologists wanted to evaluate whether it was \emph{a priori} possible to extract social map information for the Wharthon area. Zooming the map in this region, we identify 81 potentially interesting tweets (see figure \ref{fig:heatmap}b). Removing the textual filter yields approximately 3600 tweets instead. This area therefore represent an almost negligible part of our corpus. Intuitively, the tweet density is strongly tied to the population density. However, the retrieved set is small enough to allow a manual inspection, which can be useful for qualitative tests.

\section{Textual content} \label{sec:textual}

\subsection{Representation space} 

In order to categorize tweet texts, an adequate representation space has to be chosen. In the literature on event detection and sentiment analysis, the usage of chosen keywords is reported \cite{sakaki2010earthquake}, as well as classical representations such as Tf-Idf or Bag-of-Words \cite{gao_mapping_2018,kouloumpis2011twitter,batoolprecise}. These methods are based on word frequencies, and work well with structured and curated text (e.g. Wikipedia pages). However, text posted on social networks contain many abbreviations, slang and typos. Pre-processing such text has been suggested in the literature \cite{lampos_nowcasting_2012}, but this may lead to information loss or distortion.

Alternatively, we chose to use Tweet2Vec, a character-based representation space for social media content \cite{dhingra_tweet2vec:_2016}. The motivation for character-based embeddings is to be more robust to short and informal text than word-based embeddings. This method relies on extracting the final hidden state of a recurrent neural network trained to predict hashtags contained in tweets. Besides open-sourcing an implementation of their model\footnote{https://github.com/bdhingra/tweet2vec}, the authors also released pre-trained model parameters, taking 2829 characters as potential input (e.g. characters, digits, punctuation, emoticons), and returning 500-dimensional embedding vectors. This pre-trained model was used for our preliminary experiments. We collected our own corpus of similar size by sampling the Twitter stream in English language for future experiments.

\subsection{Named entity extraction}

Extracting named entities, specifically place names, can be critical for event detection. In case of natural disasters, there is an important amount of tweets mentioning affected tiers \cite{middleton_real-time_2014} (e.g. \textit{"My sister flooded Lumberton, Texas. Walmart on 69.. need help she is stranded and the family from Houston can not get to her"} in our corpus). In such situation, tweet geotags can be amended by the detected named entities. This detection can be performed by combining named entity detection libraries\footnote{https://github.com/ushahidi/geograpy} to Open Street Maps-based geocoding systems such as Nominatim\footnote{https://nominatim.openstreetmap.org/}.

In the end, as seen in section \ref{sec:experiments}, only a small portion of our corpus can be localized in a satisfactory way in the context of the targetted application. In future work we will also try to benefit from named entities present in tweet texts, e.g. \emph{"The Intersection of Asford Pkwy and Dairy Ashford Rd is singificantly higher than yesterday"}. Such an extraction has been considered in the literature about event detection in Twitter corpora \cite{middleton_real-time_2014}. Other work has exploited tags and laguage geographicity \cite{kordopatis-zilos_placing_nodate}, but such approaches would be difficult to transfer to our regional scale \emph{a priori}.

\section{Content classification} \label{sec:classification}

\subsection{Active Learning Experiments}

We define 3 target categories for tweets, already considered in the literature in a similar context \cite{bischke_multimedia_nodate}:
\begin{enumerate}
    \item \emph{Non Relevant}: the text is not related to the flood event
    \item \emph{Positive Indication}: the text is a report by a person directly affected by the flood
    \item \emph{Negative Indication}: the text is a report of a person safe from the flood
\end{enumerate}

Owing to the size of the collected corpus, and the cost of manual work that prevents exhaustive annotations, we consider active learning \cite{dawodactive} as means to reduce the corpus annotation cost.
We perform initial experiments anyway, with a sample of 421 tweets annotated by hand (316 for training and 105 for test). To favor class balance, we used the \emph{flood} keyword in our sampling process.

We then assessed how several active learning strategies behaved in the context of our classification task and representation space. These experiments were coded in Python using the \emph{libact} library \cite{yang2017libact} that implements the most common strategies.
We compared the \textit{Uncertainty sampling} to \textit{Hierarchical Sampling},with a random sampling strategy as a baseline. We monitored the precision of an SVM classifier during the execution of the strategy. The results were not seen as conclusive, as both strategies do not perform better than random. This may be due to the imbalance between the small sample size and the embedding dimensionality, thus calling for a smart way of acquiring labels.

\subsection{Crowdsourcing}

In order to train a quality classifier, we need to label our corpus as reliably and exhaustively as possible. In this view, we propose to combine active learning to the usage of a crowdsourcing platform such as \textit{Mechanical Turk} \footnote{https://www.mturk.com/} or Figure Eight \footnote{https://www.figure-eight.com/}. Our intuition is to use active learning to select the items that will be pushed to the crowdsourcing platform.

Literature shows that crowdsourcing raises reliability issues that need to be addressed \cite{nowak_how_2010}. Related work has been made in the context of image concepts annotation \cite{nowak_how_2010,loni_getting_2014}, social media analysis \cite{kamel_boulos_crowdsourcing_2011} and named entity identification \cite{finin2010annotating}. For our task, particular attention will have to be put in the way questions are asked to workers, in order to better guide them and limit the risk of errors.

\section{Conclusion}

The final objective of the project is to aggregate relevant content w.r.t. a 2D spatial grid. Spatio-temporal classification methods were proposed in the literature \cite{helwig_analyzing_2015,anantharam_extracting_2015,tamura_density-based_2013}, sometimes accounting for the out of event local tweet emission rate \cite{gao_mapping_2018} and trending topic detection \cite{atefeh_survey_2015,cordeiro_online_2016}. The present paper let us quantify the quantity and reliability of geographical information in our corpus. Besides adapting filtering methods mentionned above, we will evalute the improvement yielded by named entity extraction.

Practically, as discussed in section \ref{sec:classification}, we will explore the potential of active learning and crowdsourcing in order to identify relevant content more finely than using keywords (e.g. as made in section \ref{sec:experiments}), and harness the multimodality of the content under study (i.e. text, GPS coordinates, time, images) while accounting for potential missing values \cite{brangbour_extracting_2018}.

\section{Acknowledgements}

This work was performed in the context of the Publimape project, funded by the CORE programme of the Luxembourgish National Research Fund (FNR).

\bibliographystyle{plain}
\bibliography{refs}

\end{document}